# Broad-band plasmonic isolator compatible with low-gyrotropy magneto-optical material

Sevag Abadian*, Giovanni Magno, Vy Yam, Béatrice Dagens**

Université Paris-Saclay, CNRS, Centre de Nanosciences et de Nanotechnologies, 91120, Palaiseau, France.

*sevag.abadian@c2n.upsaclay.fr

**beatrice.dagens@c2n.upsaclay.fr

**ABSTRACT:** Optical isolator remains one of the main missing elements for photonic integrated circuits despite several decades of research. The best solutions up to now are based on transverse magneto-optical effect using either narrow-band resonators or high-gyrotropy magneto-optical materials with difficult integration on usual photonic platforms. We propose in this paper a radically new concept which enables performing broad-band non-reciprocal transmission even in the case of low-gyrotropy material. The principle explores the separation of back and forth light paths, due to the magneto-biplasmonic effect, i.e., the coupled mode asymmetry induced in plasmonic slot waveguides loaded with a magneto-optical (MO) layer. We show numerically that such a metal-MO dielectric-metal slot waveguide combined with suitable side-coupled lossy rectangular nanocavities gives more than 18 $dB$ isolation ratio on several tens of nanometers bandwidth, with only 2 $dB$ insertion losses. We propose an analytical approach describing such a magneto-plasmonic slot waveguide to identify the involved physical mechanisms and the optimization rules of the isolator. Additionally, we show that low-gyrotropy material (down to ~0.005) can be considered for isolation ratio up to 20 $dB$, opening the road to a new class of integrated isolators using easy-to-integrate hybrid or composite materials.

## INRODUCTION

The issue of the monolithic integration of optical isolators, and more generally of non-reciprocal devices in photonic circuits appeared with the rise of optical fiber telecommunications. Indeed, isolators are essential elements for the stability of laser diodes emission, and optical circulators can considerably enrich photonic circuits architecture. Their monolithic integration is the best solution to limit devices and circuits cost. Optical transmission non-reciprocity requires a propagation medium with together time and spatial symmetry breaking[1]. Over time, different mechanisms have been proposed, explored and improved to attain time-reversal breaking such as magneto-optics (MO)[2], optical nonlinear methods[3], dynamic modulation[4] or optomechanical interactions[5]. However, the most traditional way, and the most efficient one remains the use of magneto-optical effects. A first idea for integrated isolator was to reproduce the principle of bulk Faraday isolators[6,7], but the birefringence of the planar waveguides and the requirement for integrated polarizers prevent the realization of performing non-reciprocal devices. TMOKE (Transverse Magneto-Optical Effect) appeared then as the most compatible solution with planar rectangular waveguides: indeed, it is a reflection effect occurring at the MO material interface and it does not modify the light polarization. Additionally, integrated designs could be envisaged in standard photonic material systems with a MO layer as a top or side cladding. In this way two groups[8,9] simultaneously proposed to use transversely magnetized metallic MO cladding layers and exploit the imaginary part of the TMOKE, i.e., the non-reciprocal dichroic loss. Giving the whole structure an amplifying layout could compensate the remaining loss in the forward direction. The principle being inherently polarization dependent, non-reciprocal amplifier designs for TM mode[10,11] or for TE mode[12] were thus proposed and demonstrated for typically 12 $dB$ isolation at transparency. In the meantime, Fujita *et al* used the phenomena of non-reciprocal phase shift (NRPS) induced by TMOKE. A new type of optical isolators in a TM mode configuration based on a garnet waveguide Mach Zehnder Interferometer (MZI) was demonstrated[13], and considerably improved since then[14]. An important aspect of these types of isolators is that they face challenges such as complex integration on semiconductor platforms, they are much bigger in size compared to other structures and they operate at a narrow bandwidth. In search for isolators with better performance, reduced size and easier integrability, magnetic ring resonator structures have also been inspected. Since its first proposition in 2007 by N. Kono[15], several groups developed garnet oxide direct growth on semiconductor platforms[16,17], and/or highly resonant structures[18–20]. These structures can provide high isolation ratio and their miniaturized size facilitates integration with other optoelectronic devices, especially with semiconductor lasers. However, due to the nature of the ring resonator's sharp resonance, these isolators remain very sensitive to the geometrical parameters which limit their performance over a wide range of wavelengths. Alternative non-magneto-optical designs have also been proposed along these years based on different non-linearities[21–25], without improving MO garnet devices neither reducing operation budget nor footprint. MO garnet-based devices have undoubtedly led to the best device performance up to now. Nevertheless, garnet MZIs and ring resonators remain very narrow bandwidth solutions which cannot address all the applications.

In this paper, we propose a radically new guided isolator structure with respect to the mainstream of these last years. The principle of this new isolator is based on plasmonic slot waveguide modes which lose their symmetry by the use of a magnetized MO material inside the metal/dielectric/metal (MDM) slot. The phenomenon was identified a few years ago firstly in the case of anisotropic slot waveguide[26], and then more specifically in different magneto-plasmonic works[27,28]. More recently devices using this effect have been theoretically proposed for (reciprocal) switching in a plasmonic thin metallic waveguide[29] or modulation in a slot waveguide[30]. But no design was proposed for non-reciprocal devices, maybe because these structures have at first sight spatially symmetric feature.

Here we firstly detail the basic principle of the proposed guided isolator and we numerically evaluate its expected performance through relevant examples using "classical" MO material like Bismuth Iron Garnet (BIG) for operation around 1.55 $\mu m$. Then we derive an analytical model to understand the underlying physical mechanisms of the MO slot waveguide mode generation and to propose and generalize future designs compatible with fully integratable (low-gyrotropy) MO materials.

**ISOLATOR PRINCIPLE AND SPECIFIC DESIGN**

Figure 1 depicts the proposed isolator operating at telecom wavelengths, which exploits a slot (MDM) waveguide whose core dielectric material has MO properties, e.g., BIG. The fundamental guided modes in the slot waveguide result from the coupling of two SPP (Surface Plasmon Polariton) modes which can exist simultaneously at each metal-dielectric interface. Their coupling generates two "supermodes", namely, the LRSPP (Long-Range-SPP) and SRSPP (Short-Range-SPP) modes, with respectively even and odd profiles of the electric field component. The magnetization of the MO material (directed along the y-axis) and the geometrical structure are so that each SPP undergoes TMOKE. Because of TMOKE, each supermode becomes asymmetric as schematized in Figure 1b in the case of an Au-BIG-Au slot waveguide.

We call magneto-biplasmonic effect the result of TMOKE interaction in coupled plasmonic (SPP) modes. Whereas it has been considered up to now as usual magneto-plasmonic effect[30], we underline through this new name the fact that the two-step physical mechanism (mode coupling + TMOKE) radically modifies the impact of magneto-optics on guided mode propagation, as discussed later. By exciting only one of the supermodes (preferentially the low-losses LRSPP), light follows different paths in the forward and backward directions (time symmetry breaking), as shown in Figure 1b.

By introducing optical absorbers only on one side of the slot waveguide (spatial symmetry breaking), light is no more transmitted in the corresponding direction (Figure 1a,b). Any kind of lossy element can be used, such as lossy cavities, with the condition that they do not induce any back reflections. Periodicity of these elements is not required either. As a result, isolation ratio is controlled by two independent parameters: plasmonic coupled mode asymmetry and SPP absorption provided by the optical absorbers (lossy metallic cavities) for forward/backward transmission contrast.

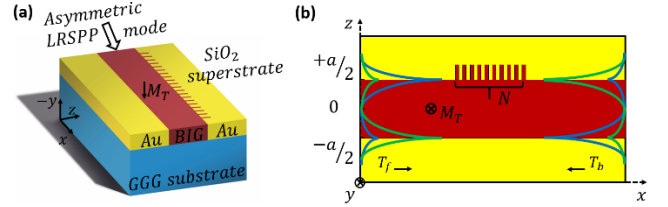

Figure 1. (a) 3D schematic view of isolator. (b) Top view of the Au-BIG-Au waveguide, with $N$ nanocavities positioned on one metal-dielectric interface. The BIG layer is magnetized along the y direction. Amplitude of electric field profiles of the LRSPP mode (green) and the SRSPP mode (blue) are represented in forward ($T_f$, increasing x) and backward ($T_b$) senses.

In this part, we first numerically evaluate the asymmetry of the 2D waveguide modes of Figure 1, before to study the lossy cavities' characteristics and finally the isolation performance of the full device. Finite Element Method (FEM) based calculations were performed using COMSOL Multiphysics 5.4, for a varying core width $a$, between 0.5 $\mu m$ and 2.5 $\mu m$, and a varying incident wavelength $\lambda_0$, between 0.7 $\mu m$ and 2.5 $\mu m$. A MDM structure thickness of $h = 0.45\ \mu m$ maximizes the optical confinement factor while maintaining vertical monomode behavior. The slot waveguide is placed on a GGG (Gallium Gadolinium Garnet) substrate with index $n_{GGG} = 1.97$ and covered with $SiO_2$ (Silica) cladding having an index $n_{SiO_2} = 1.45$.

At optical frequencies, the MO activity in this configuration results in an optical anisotropy that can be described by the following dielectric tensor, being the magnetization $\overrightarrow{M_T} = M_T \overrightarrow{u_y}$:

$$\varepsilon = \begin{pmatrix} \varepsilon_{xx} & 0 & \varepsilon_{xz} \\ 0 & \varepsilon_{yy} & 0 \\ \varepsilon_{zx} & 0 & \varepsilon_{zz} \end{pmatrix} \quad (1)$$

Having considered absorption zero ($\varepsilon'' = 0$), the diagonal elements $\varepsilon_{xx}$, $\varepsilon_{yy}$, $\varepsilon_{zz}$ are equal to the scalar isotropic dielectric constant $\varepsilon_d = 2.3^2$. The off-diagonal elements are $\varepsilon_{xz} = -\varepsilon_{zx} = ig$ where $g$ is the gyrotropy constant. $|g|$ is proportional to the magnitude of the saturation magnetization. In this part, we shall assume that $g$ is a real constant equal to $0.01^{31}$. The frequency $\omega$ dependent dielectric constant of gold is modeled by the Drude-Lorentz model fitting of ellipsometric data with plasma frequency $\omega_p = 1.29 \times$

$10^{16}\ rad/s$, plasma collision frequency $\gamma_c = 6.47 \times 10^{13}\ rad/s$ and $\varepsilon_\infty = 1$ the high-frequency contribution to the relative permittivity[32]:

$$\varepsilon_m(\omega) = \varepsilon_\infty - \frac{\omega_p^2}{\omega(i\gamma_c + \omega)} \qquad (2)$$

The mode asymmetry percentage is one of the key parameters of the isolator design: it should be maximized versus geometrical dimensions to optimize the isolation ratio. Two-dimensional modal analysis is performed on the proposed slot waveguide isolator to calculate LRSPP modes at the entrance of the device. The mode asymmetry is deduced from the absolute relative difference of the magnetic field intensities at both metal-dielectric interfaces.

As shown in Figure 2, asymmetry is strongly dependent on geometry and on wavelength: for a fixed wavelength, the increasing width of the dielectric waveguide enhances the asymmetry of the LRSPP mode; besides, the asymmetry is also higher for lower wavelengths. These first data indicate that the isolation ratio could be geometrically controlled.

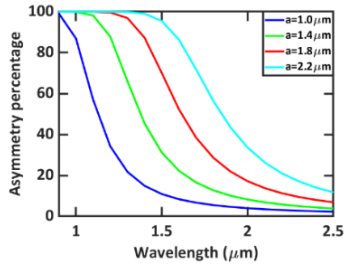

Figure 2. Electromagnetic field asymmetry percentage in Au-BIG-Au slot waveguide versus the wavelength and for dielectric (BIG) widths $a = 1.0\ \mu m$, $1.4\ \mu m$, $1.8\ \mu m$ and $2.2\ \mu m$ for a fixed slot waveguide (BIG and Au) thickness $h = 0.45\ \mu m$.

Besides, unlike previously proposed TMOKE isolators, propagating modes have identical effective indices in the forward and backward directions for waveguide symmetry reasons since the magnetization is applied along the y-axis; however light energy follows different paths (opposite interfaces of the MDM), and the one-side lossy nanocavities provide absorption on the backward signal. Although many complex nanocavity geometries could be considered, we suggest here elementary rectangular nanocavities as represented in Figure 3, which are easy to integrate in the isolator structure with a manageable nano-fabrication process. The geometrical parameters of these recurrently placed nanocavities such as the length, $l_{cav}$ (in the x direction) and width $w_{cav}$ (in the z direction) can be properly and specifically designed to obtain the required optical response (high losses on a large bandwidth). Two other parameters which need to be addressed are the periodicity $p_{cav}$ and the number of cavities $N$. Periodicity should be adequately chosen to avoid both the coupling between adjacent cavities and back reflections, whereas the number $N$ should be as low as possible to decrease the total isolator length and minimize the plasmonic losses in the forward direction, and sufficiently high to provide complete absorption of the backward propagating light.

The resonant condition of the rectangular nanocavity can be approximated by the Fabry-Perot (FP) equation[33]:

$$2m\pi = 2n_{eff,g(x,y)}k_0 w_{cav} + \varphi_{r1} + \varphi_{r3}, m \in \mathbb{Z} \qquad (3)$$

where $m$ is the order of the horizontal FP resonance and $n_{eff,g(x,y)}$ is the effective index of the fundamental mode within the BIG gap of the unitary cell calculated by considering a cross section parallel to the (x,y) plane. $\varphi_{ri}$ is the reflection phase shift of the gap mode at each end of the cavity, with $i = 1$ and $i = 3$ corresponding to the reflection of light respectively from the BIG waveguide and from the Au-surrounding.

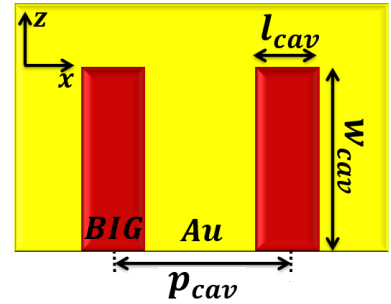

Figure 3. Schematic of the lossy rectangular cavity formed by BIG and surrounded by Au.

In Figure 4, FDTD calculations give the absorption spectra for a fixed cavity length $l_{cav} = 0.1\ \mu m$ and different $w_{cav}$. From the FP equation it is quite evident that the increase in cavity width $w_{cav}$ results in a redshift of a given FP resonance. Higher order horizontal FP resonant modes are also expected to appear as a result of this increase: for this reason, as $w_{cav}$ grows the cavity absorption peak becomes sharper and stronger. Thus, a rectangular cavity with dimensions $l_{cav} = 0.1\ \mu m$ and $w_{cav} = 0.95\ \mu m$ can provide an absorption peak near $\lambda_0 = 1.55\ \mu m$ due to the fourth order horizontal Fabry Perot resonance happening inside the cavity as shown in the magnetic field plot of Figure 4e. Several cavities with these dimensions, integrated adjacent to each other on one interface of the MDM structure, are capable of absorbing the SPPs propagating on its side.

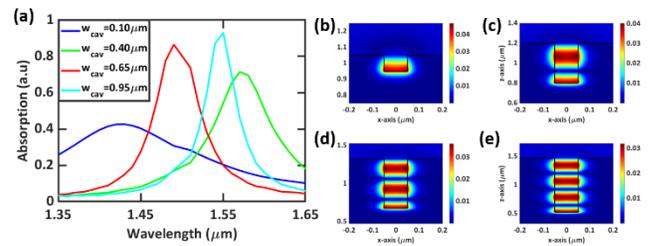

Figure 4. Absorption spectra of a rectangular nanocavity as a function of $\lambda_0$ and $w_{cav}$ for cavity length $w_{cav} = 0.10\ \mu m, 0.40\ \mu m,$

$0.65\ \mu m$ and $0.95\ \mu m$. (b-e) Intensity of the magnetic field along the (x,z) plane in increasing order presented in (a).

At this point, all the necessary components required to build a functioning isolator as shown in Figure 1 have been discussed above. Using commercial software Lumerical based on FDTD method, 3D simulations have been performed in order to simulate the propagation of light in the forward and backward directions of the full structure. Isolation ratio, which is a quantitative measurement of the difference between the forward to the backward transmissions in $dB$, is defined as follows:

$$IR\ (dB) = 10\ log_{10}(T^f/T^b) \qquad (4)$$

Another relevant parameter is the Figure of Merit (FoM), which is the ratio of isolation to the insertion losses. FoM expresses the efficiency of the isolator structure and can be represented by the equation:

$$FoM = IR\ (dB)/10\ log_{10}(T_f) \qquad (5)$$

In order to obtain optimum results for propagation simulations, a wide waveguide width $a = 2.4\ \mu m$ which promises a high asymmetry percentage is chosen to obtain the highest contrast between forward and backward transmissions in the wavelength range of $1.425\ \mu m$ to $1.675\ \mu m$. In the coming results, rectangular cavities have dimensions $l_{cav} = 0.1\ \mu m$ and $w_{cav} = 0.95\ \mu m$. Although the cavity arrangement is not of primary importance and randomly placed cavities can provide similar intended SPP absorption, we have positioned the ten cavities with a regular spacing $p_{cav} = 0.8\ \mu m$, which is sufficient to avoid their mutual coupling and allows each cavity to absorb independently from the others.

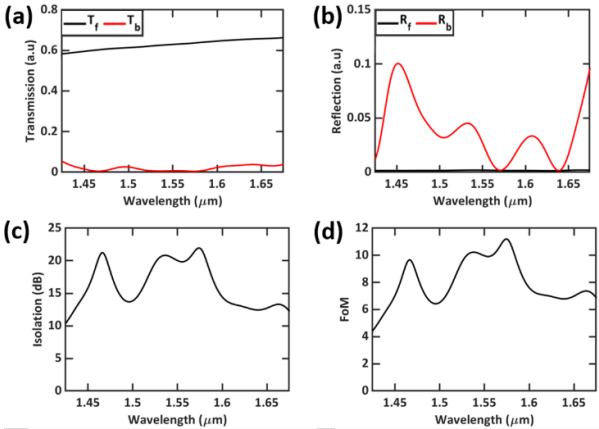

Figure 5. (a) Forward and backward transmission $(T_f, T_b)$ and (b) reflections $(R_f, R_b)$. (c) Isolation ratio and (d) FoM for an isolator with width $a = 2.4\ \mu m$, height $h = 0.45\ \mu m$ and total length of $11.2\ \mu m$.

In the forward direction the transmission curve takes a smooth shape (black curve in Figure 5a) while the majority of the LRSPP energy travel on one interface of the slot waveguide with very small interaction with the cavities as shown by the amplitude of the magnetic field surface plot in Figure 6a. Negligible backward reflections occur in this case (black curve in Figure 5b). On the other hand, the backward propagation of the LRSPP energy occurs on the interface with the periodically arranged nanocavities which provide absorption and induce a very perturbed and low transmission (red curve in Figure 5a), this corresponds to the magnetic field plot of Figure 6b. In this sense of propagation some reflections occur due to the interaction of the SPP with the cavities (red curve in Figure 5b). These reflections have moderate impact on the isolator performance since they induce propagation in the increasing x direction. As we can notice in Figure 5c,d high isolation ratio and FoM are obtained over the wavelength range of interest which are due to i) the high asymmetry percentage of the coupled LRSPP mode hosted by the magnetized MO slot waveguide, and ii) the losses due to horizontal FP resonances of the rectangular cavities. Notably, isolation ratio higher than $18\ dB$ and FoM higher than 9 are obtained between wavelengths $1.52\ \mu m$ and $1.58\ \mu m$, with the highest value of isolation ratio reaching $22\ dB$ and FoM 11 around $\lambda_0 = 1.57\ \mu m$.

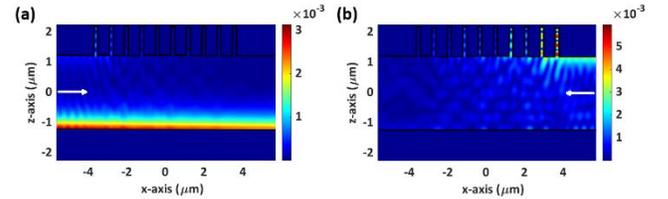

Figure 6. Intensity of magnetic field along the (x,z) plane (top view of the device), when $y = 0\ \mu m$ and $\lambda_0 = 1.57\ \mu m$, for (a) forward and (b) backward transmission directions respectively, in the case of isolator with width $a = 2.4\ \mu m$, height $h = 0.45\ \mu m$ and total length of $11.2\ \mu m$.

The simple design we proposed (consisting of a single linear array of absorbing cavities) already shows the potential of this non-reciprocal structure. It is worth underlining that the different degrees of freedom shown by the structure enable a high isolation value with both low back reflections and high FoM values over a large spectral range.

## ANALYTICAL DERIVATION, TOWARD LOW-GYROTROPY DESIGN

The following analytical framework is developed in order to deepen the physical understanding of the magneto-biplasmonics effect in MO slot waveguides and to identify design optimization rules especially in the case of more integratable low-gyrotropy MO materials. We start our study by deriving an analytical model based on Maxwell's equations applied to the magneto-optical MDM structure shown in Figure 1b. The (coupled) SPP modes in the slot waveguide are composed of the three non-zero field components $E_x$, $H_y$ and $E_z$, where x is the propagation direction, y the magnetization direction, and z the transverse direction. These electromagnetic fields have a maximum value at each dielectric-metal interfaces and exponentially decay in both the metal and the dielectric material. By using Maxwell's equations in magneto-optical media, i.e., considering the MO dielectric tensor including off-diagonal elements as shown in eq 1, the wave equation and the relation between the electromagnetic fields can be derived. The solutions for the magnetic field y-component $H_y$ that satisfy the $H$-field boundary conditions at $z = \pm a/2$, where $a$ is the width of the dielectric region inside the MDM, can be written as:

$$e^{i\beta x} \begin{cases} Ae^{-k_m(z-a/2)}, & z \geq +a/2 \\ Ce^{k_d(z-a/2)} + De^{-k_d(z+a/2)}, & -a/2 \leq z \leq +a/2 \\ Be^{k_m(z+a/2)}, & z \leq -a/2 \end{cases} \quad (6)$$

where $k_m$ and $k_d$ are the wavevectors z-components in the metal and dielectric layers respectively and $\beta$ is the propagation constant of the traveling wave in the x direction.

The dispersion relation of the optical modes supported by the MDM structure, with MO material having the diagonal element of the dielectric tensor $\varepsilon_d = n_{MO}^2$, can be derived by applying the boundary conditions on the tangential electromagnetic fields at the metal-dielectric interfaces (eq 6):

$$e^{-2k_d a} = \frac{(\beta \gamma_{xz}^d)^2 + (k_d \gamma_{xx}^d + k_m \gamma_{xx}^m)^2}{(\beta \gamma_{xz}^d)^2 + (k_d \gamma_{xx}^d - k_m \gamma_{xx}^m)^2} \quad (7)$$

where $\gamma_{xz}^d = ig/(n_{MO}^4 - g^2)$, $\gamma_{xx}^d = n_{MO}^2/(n_{MO}^4 - g^2)$, $\gamma_{xx}^m = 1/n_{Au}^2$ and the wavevectors $k_m$ and $k_d$ can be represented by $k_i = \sqrt{\beta^2 - k_0^2/\gamma_{xx}^i}$, $i = m,d$ and $k_0 = 2\pi/\lambda_0$, $k_0$ being the wavenumber in free space and $\lambda_0$ the operating optical wavelength.

By considering $g^2 \ll n_{MO}^4$ and $\varepsilon_m - \varepsilon_d \cong \varepsilon_m$ the complex propagation constant $\beta$ of the MDM waveguide plasmonic modes, LRSPP and SRSPP, can be formalized and expressed in a comprehensible way:

$$\beta \approx \beta_0 \pm \sqrt{\beta_g^2 + \beta_a^2} \quad (8)$$

where $\beta_g = \frac{\beta_0 g}{(1-\varepsilon_d^2/\varepsilon_m^2)\sqrt{-\varepsilon_d \varepsilon_m}}$, $\beta_a = \frac{2\beta_0 e^{-k_d a}}{(1-\varepsilon_d^2/\varepsilon_m^2)}\left(\frac{-\varepsilon_d}{\varepsilon_m}\right)$ and $\beta_0 = k_0 \sqrt{\frac{\varepsilon_m \varepsilon_d}{\varepsilon_m + \varepsilon_d}}$ is the wavenumber of an SPP on a single metal-dielectric interface without magnetization. The "plus" and "minus" signs in eq 8 correspond respectively to the LRSPP and SRSPP modes.

$\beta_g$ and $\beta_a$ can be seen as modifications to the single SPP propagation constant $\beta_0$. These two terms are respectively related to the gyrotropy and to the SPP modes coupling (strength and overlap) of the magneto-optical slot waveguide. If an infinite thickness a is considered (which corresponds to $\beta_a \approx 0$), eq 8 gives the expression of the single MO SPP propagation constants, depending on the considered interface (or equivalently on propagation direction, or on g sign[34]). On the contrary if $g = 0$ ($\beta_g \approx 0$), the same equation reduces to the canonical non-MO LRSPP and SRSPP propagation constants. For application purposes, it is much more convenient to consider the LRSPP mode only because of lower propagation losses with respect to the SRSPP mode.

It is worth noting that the dispersion relation represented in eq 7 is reciprocal since it depends on the square of the propagation constant $\beta^2$. Thus, the numerical variation observed in $\beta$ (or as a result in the mode effective index $n_{eff}$) is dependent solely on gyrotropy absolute value $|g|$, on slot width $a$ and on material permittivities, but not on the propagation direction.

In the non-magnetic case ($\beta_g = 0$) the amplitude of the magnetic field $H_y$ is the same at the two dielectric-metal interfaces ($A = B$). However, and upon the presence of magnetization in the y direction (time symmetry breaking) in the sandwiched dielectric layer, the electromagnetic field distribution is altered and the SPP amplitude at the two dielectric-metal interfaces are no longer identical: $A \neq B$. In order to quantify the degree of the mode asymmetry, the parameter $\phi$ is defined as the percentage of the normalized absolute difference between the amplitude of the magnetic field at $z = +a/2$ and at $z = -a/2$. $\phi$ can be deduced using eqs. 6 and 7.

$$\phi = 100 \times \left(1 - \left|\frac{H_y(z=+a/2)}{H_y(z=-a/2)}\right|\right) \approx 100 \times \left(1 - \left|\sqrt{\frac{1-\alpha}{1+\alpha}}\right|\right) \quad (9)$$

where $\alpha = g/(2e^{-k_d a}\varepsilon_d^{3/2}/\sqrt{-\varepsilon_m})$

As it can be seen from the derived eq 9 and the complex expression of $\alpha$, the quantity $\phi$ is in direct relation with the gyrotropy of the magneto-optical material $g$ (including its sign), the waveguide width $a$ and the incident wavelength $\lambda_0$. For a given metal and at a fixed wavelength, the width $a$ and the permittivity (diagonal element) of the magneto-optical layer have the strongest influence on the mode asymmetry, since they both contribute to the exponential term. For our application, both the width and the MO material dielectric constant should thus be increased, in the limit of the existence of the coupled LRSPP mode. This limit can be assessed by the mode losses which are lower for LRSPP mode than the uncoupled SPP one.

The derived analytical model can be used now to evaluate design rules and expected performance in term of mode asymmetry. To first check the validity of this approximated model in Figure 7, we have calculated the degree of mode asymmetry $\phi$ by using the analytical model and COMSOL, in the case of two values of gyrotropy (0.1 and 0.01), for arbitrary theoretical dielectric material permittivities (diagonal element). The overall tendency is well reproduced by the analytical model.

According to Figure 7a,b the decrease in asymmetry due to the decrease in gyrotropy $g$ can be compensated by the increase in the refractive index of the magneto-optical dielectric. Nevertheless, the impact of refractive index is more strongly marked when gyrotropy is low. For a 2 $\mu m$ waveguide width and an optical index equal to 3, the asymmetry of the LRSPP mode can reach more than 95 % in the case of very high gyrotropy 0.1, and more than 80 % in the case of the more realistic value $g = 0.01$.

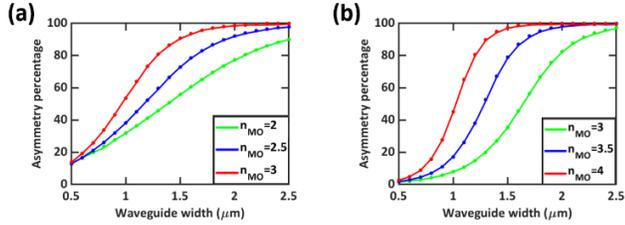

Figure 7. (1D case) Magnetic field asymmetry percentage for gyrotropy (a) $g = 0.1$, and $n_{MO} = 2, 2.5$ and 3; (b), $g = 0.01$ and $n_{MO} = 3, 3.5$ and 4; for a fixed incident wavelength $\lambda_0 = 1.55\ \mu m$ and waveguide width ranging between $0.5\ \mu m$ and $2.5\ \mu m$. Straight lines represent results obtained from modal study using COMSOL Multiphysics whereas the asterisks represent results from analytical model.

We can also consider a new class of composite MO materials with gyrotropy as low as $g = 0.005$ which have been recently demonstrated[35]. In such a composite material the refractive index can be managed partly independently of the MO properties. Even if the gyrotropy is quite low in such a diluted material, the effect on asymmetry of the coupled mode can be compensated by the effective index. In this context we have considered in Figure 8 the expected asymmetry induced in magneto-biplasmonic waveguides with different refractive indices, in the case of $g = 0.005$.

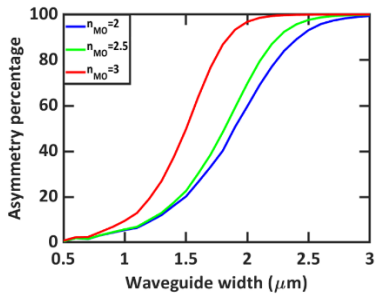

Figure 8. (2D case) Electromagnetic field asymmetry percentage in Au-BIG-Au slot waveguide for gyrotropy $g = 0.005$, and a variation of the dielectric refractive index $n_{MO} = 2, 2.5$, and 3, for a fixed incident wavelength $\lambda_0 = 1.55\ \mu m$, waveguide width $1.2\ \mu m$ and thickness $h = 0.45\ \mu m$.

Promising results show asymmetry higher than 95 % with realistic integratable MO material ($2 < n_{MO} < 2.5$). In all the cases, the simplest parameter to tune remains the waveguide width[36].

**DISCUSSION**

We have shown through specific examples the potential of magneto-biplasmonics for optical isolators. Nevertheless, the analytical model can be further analyzed in order to show that this concept is quite general. Indeed, eq 8 gives similar expressions as the eigenvalues obtained with the Improved Coupled-Mode Theory[37] in the case of the coupling of two different waveguides: the both coupled waveguides would be here the interfaces supporting the SPPs, their "difference" being induced by the gyrotropy (via TMOKE), and their coupling factor would be related to the SPP penetration into the dielectric waveguide, included in the term $e^{-k_d a}$. Additionally it appears that $\alpha$ (in eq 9) is equal to the ratio $\beta_g / \beta_a$, i.e., the asynchronism factor of the coupled modes as defined in Ref[37]. This asynchronism factor represents the competitive contribution to the coupled mode asymmetrisation, of the both SPPs 'difference' (indicated by $\beta_g$ which favours the asymmetrisation) and of their coupling strength and/or overlap (indicated by $\beta_a$ which counteracts it). The plasmonic nature of the both coupled 'waveguides' (interfaces) makes here the asynchronism factor very sensitive to geometrical parameters: indeed the mode coupling decreases rapidly with the increase of slot waveguide width whereas the optical intensity remains concentrated at the MO-metal interface where TMOKE occurs. Asymmetry increases thus with width $a$. Also for this reason, higher asymmetry is obtained for lower wavelengths, at which more confined SPP can be obtained as well for gold (Figure 2) as for silver slot waveguides. Additionally, the light concentration at the plasmonic interfaces is also enhanced by the dielectric permittivity which has strong impact on mode profile through the term $e^{-k_d a}$. We expect similar asymmetry of the supermodes in coupled MO all-dielectric waveguides system, without the advantage of the light concentration and of the TMOKE enhancement induced by plasmonic effect. Magneto-biplasmonics provides thus a technological breakthrough in the integrated non-reciprocal devices domain, which could be combined with magneto-biphotonic (nonmetallic) structures for lower insertion losses.

Besides such behavior is not limited to MDM structures neither to non-reciprocal devices. Similar (but different) formulations are obtained for DMD switches[29], and several theoretical works suggest that the MO properties could be present as well in the central as in the sides layers of the slot waveguide[28–30]. This opens the route to many innovative non-reciprocal devices.

**CONCLUSIONS**

For decades, the performance of integrated photonic isolators was limited by integration issues, high insertion losses or low operation bandwidth. Here we propose a novel isolator geometry based on an MDM waveguide configuration with a newly exploited phenomena "the magneto-biplasmonic" effect which has the potential to tackle previous issues. Time reversal symmetry is broken by the MO properties of the waveguide central layer, and spatial symmetry is broken by lossy cavities, to finally achieve non-reciprocal transmission of SPPs. It should be noted that the rectangular absorbing cavities can be interchanged with different geometries as long as SPP absorption at the wavelength of interest is realized. We have validated that by using Maxwell's equations in gyrotropic media with an asymmetric tensor a 1D model is sufficient enough to predict and give a good estimation of the 2D results (mainly asymmetry percentage) obtained by the FEM method. Moreover, we have proved analytically and numerically that the geometrical parameters as well as material properties (dielectric constant and gyrotropy) of the MDM slot waveguide can be chosen according to the incident wavelength range of operation, and that they are all relevant parameters in determining the final isolation performance.

Compared to former designs, the main issues tackled are the low operation bandwidth and the high insertion losses. Our proposed new generation of photonic isolators is tunable and can be easily integrated with silicon strip waveguides operating with the fundamental TE mode. By following appropriate device engineering, that is to say impedance matching, an efficient coupling design can be defined which could excite solely the asymmetric LRSPP mode with minor back reflections. The remaining challenge is to realize this transition in a short propagation distance.

Finally, by this paper we underline the importance of the "biplasmonic" effect of the magneto-plasmonics field as the most efficient tool for achieving optical signal isolation up to $20 \sim 30\ dB$ on a broad band and by using easy-to-integrated low gyrotropy MO materials, and emphasize the exciting properties of magneto-optical slot waveguide configuration for future photonic circuit components.

## METHOD

Finite Element Method (FEM) based calculations were performed using COMSOL Multiphysics 5.4 including the Electromagnetic Waves Frequency Domain module (emw) physics interface. The COMSOL simulation area is composed of a three-layer heterostructure identical to the MDM geometry represented in Figure 1a. Extremely fine triangular mesh is used, with largest element size $\approx a/100$, at the dielectric-metal boundaries to give reliable and persistent results. The structure is excited from the left with a Numeric port, while scattering boundary condition is applied on the other boundaries of the structure to eliminate any back reflections that could alter the results. Using Boundary Mode Analysis for the input port, the properties of interest of the 1D MDM waveguide can be retrieved. Similarly, the properties of the 2D slot waveguide interface are obtained using modal analysis. Finally, to evaluate the mode asymmetry percentage, operators are assigned on each of the metal-dielectric interfaces to collect the electromagnetic field intensity.

Using the commercial software Lumerical FDTD Solutions two simulation tasks are performed. In the FDTD method, a spatial discretization mesh of size $\Delta x = \Delta z = 10\ nm$ is imposed on the structure which is found enough for convergence of the numerical results. First, in order to study the optical response of the unitary cell with respect to the $w_{cav}$ and $l_{cav}$, 2D simulations have been performed. A plane wave source with a frequency monitor are included in the simulation region and placed above the nanocavity to excite and record the intensity of absorption, respectively. Finally, to evaluate the performance of the isolator design., 3D simulations are launched, and the structure is excited using ports, which can simultaneously act as an exciting source for the required mode and a frequency monitor to collect transmission and reflection coefficients.

## AUTHOR INFORMATION

### Corresponding Author
* E-mail: sevag.abadian@c2n.upsaclay.fr
* E-mail: beatrice.dagens@c2n.upsaclay.fr

### Author Contributions
The manuscript was written through contributions of all authors.

## ACKNOWLEDGMENT
The authors declare no competing financial interest.

## ABBREVIATIONS

MO, magnetooptical; MDM, Metal-Dielectric-Metal; DMD, Dielectric-Metal-Dielectric; SPP, Surface Plasmon Polaritons; LRSPP, Long Range Surface Plasmon Polaritons; SRSPP, Short Range Surface Plasmon Polaritons; TMOKE, Transverse Magnetooptic Kerr Effect; NRPS, Non-reciprocal phase shift; MZI, Mach Zehnder Interferometer; FP, Fabry Perot; FoM, Figure of Merit; FDTD, Finite Difference Time Domain; FEM, Finite Element Method; GGG, Gallium Gadolinium Garnet; BIG, Bismuth Iron Garnet.

Insert Table of Contents artwork here

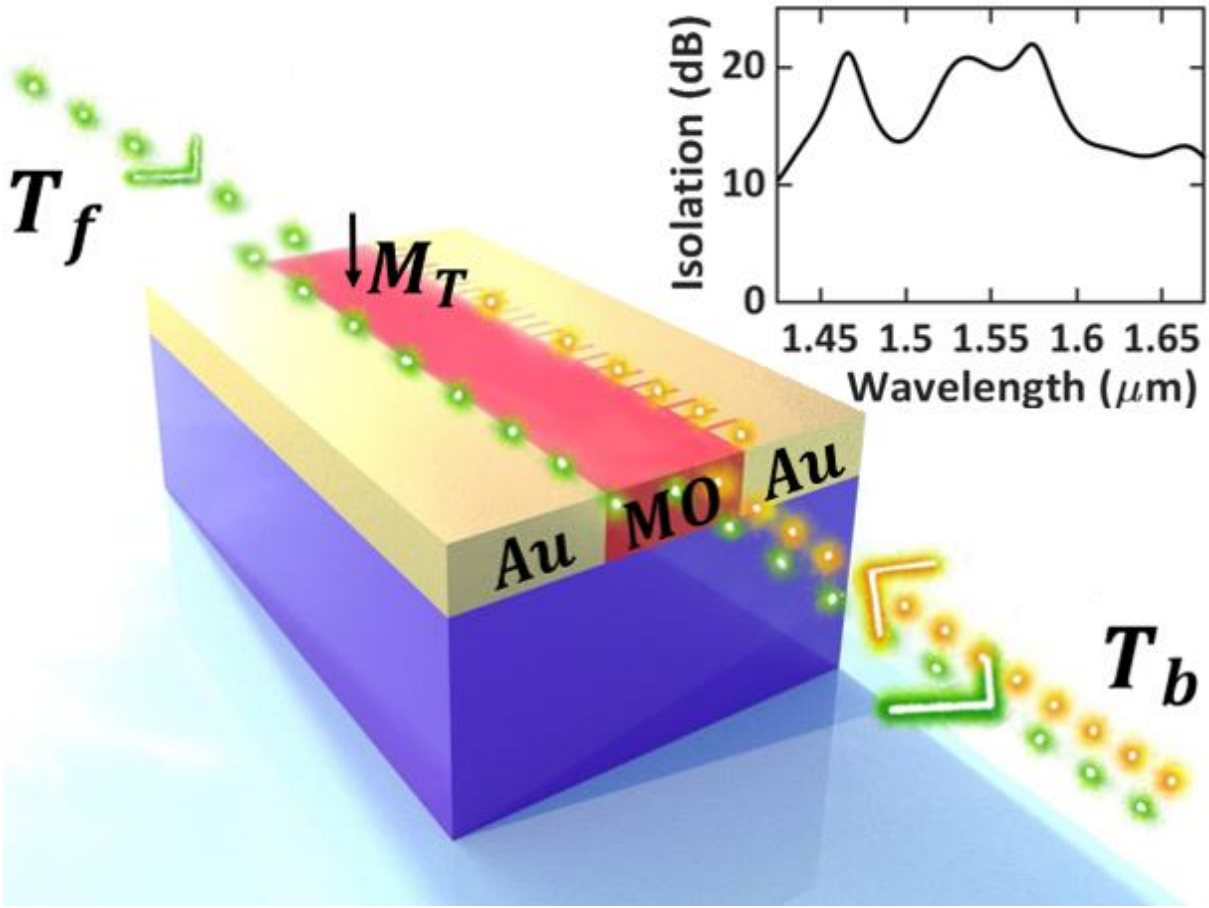